\documentstyle[twoside,psfig]{article}

\catcode`\@=11
\long\def\@makefntext#1{
\protect\noindent \hbox to 3.2pt {\hskip-.9pt  
$^{{\eightrm\@thefnmark}}$\hfil}#1\hfill}               %CAN BE USED 

\def\thefootnote{\fnsymbol{footnote}}
\def\@makefnmark{\hbox to 0pt{$^{\@thefnmark}$\hss}}    %ORIGINAL 
        
\def\ps@myheadings{\let\@mkboth\@gobbletwo
\def\@oddhead{\hbox{}
\rightmark\hfil\eightrm\thepage}   
\def\@oddfoot{}\def\@evenhead{\eightrm\thepage\hfil
\leftmark\hbox{}}\def\@evenfoot{}
\def\sectionmark##1{}\def\subsectionmark##1{}}

\oddsidemargin=\evensidemargin
\renewcommand{\thefootnote}{\fnsymbol{footnote}}

\newcounter{sectionc}\newcounter{subsectionc}\newcounter{subsubsectionc}
\renewcommand{\section}[1] {\vspace{12pt}\addtocounter{sectionc}{1} 
\setcounter{subsectionc}{0}\setcounter{subsubsectionc}{0}\noindent 
        {\tenbf\thesectionc. #1}\par\vspace{5pt}}
\renewcommand{\subsection}[1] {\vspace{12pt}\addtocounter{subsectionc}{1} 
        \setcounter{subsubsectionc}{0}\noindent 
        {\bf\thesectionc.\thesubsectionc. {\kern1pt \bfit #1}}\par\vspace{5pt}}
\renewcommand{\subsubsection}[1] {\vspace{12pt}\addtocounter{subsubsectionc}{1}
        \noindent{\tenrm\thesectionc.\thesubsectionc.\thesubsubsectionc.
        {\kern1pt \tenit #1}}\par\vspace{5pt}}
\newcommand{\nonumsection}[1] {\vspace{12pt}\noindent{\tenbf #1}
        \par\vspace{5pt}}

\newcounter{appendixc}
\newcounter{subappendixc}[appendixc]
\newcounter{subsubappendixc}[subappendixc]
\renewcommand{\thesubappendixc}{\Alph{appendixc}.\arabic{subappendixc}}
\renewcommand{\thesubsubappendixc}
        {\Alph{appendixc}.\arabic{subappendixc}.\arabic{subsubappendixc}}

\renewcommand{\appendix}[1] {\vspace{12pt}
        \refstepcounter{appendixc}
        \setcounter{figure}{0}
        \setcounter{table}{0}
        \setcounter{lemma}{0}
        \setcounter{theorem}{0}
        \setcounter{corollary}{0}
        \setcounter{definition}{0}
        \setcounter{equation}{0}
        \renewcommand{\thefigure}{\Alph{appendixc}.\arabic{figure}}
        \renewcommand{\thetable}{\Alph{appendixc}.\arabic{table}}
        \renewcommand{\theappendixc}{\Alph{appendixc}}
        \renewcommand{\thelemma}{\Alph{appendixc}.\arabic{lemma}}
        \renewcommand{\thetheorem}{\Alph{appendixc}.\arabic{theorem}}
        \renewcommand{\thedefinition}{\Alph{appendixc}.\arabic{definition}}
        \renewcommand{\thecorollary}{\Alph{appendixc}.\arabic{corollary}}
        \renewcommand{\theequation}{\Alph{appendixc}.\arabic{equation}}
        \noindent{\tenbf Appendix \theappendixc #1}\par\vspace{5pt}}
\newcommand{\subappendix}[1] {\vspace{12pt}
        \refstepcounter{subappendixc}
        \noindent{\bf Appendix \thesubappendixc. {\kern1pt \bfit #1}}
        \par\vspace{5pt}}
\newcommand{\subsubappendix}[1] {\vspace{12pt}
        \refstepcounter{subsubappendixc}
        \noindent{\rm Appendix \thesubsubappendixc. {\kern1pt \tenit #1}}
        \par\vspace{5pt}}

\newcommand{\textlineskip}{\baselineskip=13pt}
\newcommand{\smalllineskip}{\baselineskip=10pt}

\def\eightcirc{
\begin{picture}(0,0)
\put(4.4,1.8){\circle{6.5}}
\end{picture}}
\def\eightcopyright{\eightcirc\kern2.7pt\hbox{\eightrm c}}

\def\abstracts#1#2#3{{
        \centering{\begin{minipage}{4.5in}\footnotesize\baselineskip=10pt
        \parindent=0pt #1\par 
        \parindent=15pt #2\par
        \parindent=15pt #3
        \end{minipage}}\par}} 

\def\keywords#1{{
        \centering{\begin{minipage}{4.5in}\footnotesize\baselineskip=10pt
        {\footnotesize\it Keywords}\/: #1
         \end{minipage}}\par}}

\newcommand{\bibit}{\nineit}
\newcommand{\bibbf}{\ninebf}
\renewenvironment{thebibliography}[1]
        {\frenchspacing
         \ninerm\baselineskip=11pt
         \begin{list}{\arabic{enumi}.}
        {\usecounter{enumi}\setlength{\parsep}{0pt}     
         \setlength{\leftmargin 12.7pt}{\rightmargin 0pt} %FOR 1--9 ITEMS
         \setlength{\itemsep}{0pt} \settowidth
        {\labelwidth}{#1.}\sloppy}}{\end{list}}

\newcounter{itemlistc}
\newcounter{romanlistc}
\newcounter{alphlistc}
\newcounter{arabiclistc}

\newcommand{\fcaption}[1]{
        \refstepcounter{figure}
        \setbox\@tempboxa = \hbox{\footnotesize Fig.~\thefigure. #1}
        \ifdim \wd\@tempboxa > 5in
           {\begin{center}
        \parbox{5in}{\footnotesize\smalllineskip Fig.~\thefigure. #1}
            \end{center}}
        \else
             {\begin{center}
             {\footnotesize Fig.~\thefigure. #1}
              \end{center}}
        \fi}

\newcommand{\tcaption}[1]{
        \refstepcounter{table}
        \setbox\@tempboxa = \hbox{\footnotesize Table~\thetable. #1}
        \ifdim \wd\@tempboxa > 5in
           {\begin{center}
        \parbox{5in}{\footnotesize\smalllineskip Table~\thetable. #1}
            \end{center}}
        \else
             {\begin{center}
             {\footnotesize Table~\thetable. #1}
              \end{center}}
        \fi}

\def\@citex[#1]#2{\if@filesw\immediate\write\@auxout
        {\string\citation{#2}}\fi
\def\@citea{}\@cite{\@for\@citeb:=#2\do
        {\@citea\def\@citea{,}\@ifundefined
        {b@\@citeb}{{\bf ?}\@warning
        {Citation `\@citeb' on page \thepage \space undefined}}
        {\csname b@\@citeb\endcsname}}}{#1}}

\newif\if@cghi
\def\cite{\@cghitrue\@ifnextchar [{\@tempswatrue
        \@citex}{\@tempswafalse\@citex[]}}
\def\citelow{\@cghifalse\@ifnextchar [{\@tempswatrue
        \@citex}{\@tempswafalse\@citex[]}}
\def\@cite#1#2{{$\null^{#1}$\if@tempswa\typeout
        {IJCGA warning: optional citation argument 
        ignored: `#2'} \fi}}

\def\pmb#1{\setbox0=\hbox{#1}
        \kern-.025em\copy0\kern-\wd0
        \kern.05em\copy0\kern-\wd0
        \kern-.025em\raise.0433em\box0}

\def\fnt#1#2{\footnotetext{\kern-.3em
        {$^{\mbox{\scriptsize #1}}$}{#2}}}

\def\fpage#1{\begingroup
\voffset=.3in
\thispagestyle{empty}\begin{table}[b]\centerline{\footnotesize #1}
        \end{table}\endgroup}

\def\runninghead#1#2{\pagestyle{myheadings}
\markboth{{\protect\footnotesize\it{\quad #1}}\hfill}
{\hfill{\protect\footnotesize\it{#2\quad}}}}
\font\tenrm=cmr10
\font\tenit=cmti10 
\font\tenbf=cmbx10
\font\bfit=cmbxti10 at 10pt
\font\ninerm=cmr9
\font\nineit=cmti9
\font\ninebf=cmbx9
\font\eightrm=cmr8

\def\wt{\widetilde}
\def\ba{\begin{eqnarray}}
\def\ea{\end{eqnarray}}
\def\bea{\begin{eqnarray}}
\def\eea{\end{eqnarray}}
\def\qed{\hbox{${\vcenter{\vbox{                        %HOLLOW SQUARE
   \hrule height 0.4pt\hbox{\vrule width 0.4pt height 6pt
   \kern5pt\vrule width 0.4pt}\hrule height 0.4pt}}}$}}

\renewcommand{\thefootnote}{\fnsymbol{footnote}}        %USE SYMBOLIC FOOTNOTE

\begin{document}
\setlength{\textheight}{7.7truein}  %for 2nd page onwards

\runninghead{J. F. Cari\~nena \& A. Ramos}{The partnership of potentials
in Quantum Mechanics and Shape Invariance}

\normalsize\textlineskip
\thispagestyle{empty}
\setcounter{page}{1}

%\copyrightheading{}                     
%{Vol. 0, No.0 (1992) 000--000}

\vspace*{0.48truein}

\fpage{1}
\centerline{\bf THE PARTNERSHIP OF POTENTIALS IN QUANTUM}
\baselineskip=13pt
\centerline{\bf MECHANICS AND SHAPE INVARIANCE\footnote{To appear in Modern Physics Letters A}}
\vspace*{0.37truein}
\centerline{\footnotesize JOS\'E F. CARI\~NENA\footnote{E-mail: jfc@posta.unizar.es}
\ and  ARTURO RAMOS\footnote{E-mail: arrg@posta.unizar.es}}
\baselineskip=12pt
\centerline{\footnotesize\it Departamento de F\'{\i}sica Te\'orica, 
Facultad de Ciencias,}
\baselineskip=10pt
\centerline{\footnotesize\it Universidad de Zaragoza, 50009--Zaragoza, Spain}
\vspace*{5pt}

%\publisher{(received date)}{(revised date)}

\vspace*{0.21truein}
\abstracts{The concept of partnership of potentials is studied in detail
and in particular the non--uniqueness due to 
the ambiguity in the election of the factorization
energy and in the choice of the solution of certain Riccati equation.
We generate new factorizations from old ones using 
invariance under parameter transformations. 
The theory is illustrated with some examples.
}{}{}

\vspace*{10pt}

%\textlineskip                  %) USE THIS MEASUREMENT WHEN THERE IS
%\vspace*{12pt}                 %) NO SECTION HEADING

\vspace*{1pt}\textlineskip      %) USE THIS MEASUREMENT WHEN THERE IS
\section{Introduction}          %) A SECTION HEADING
\vspace*{-0.5pt}
\noindent
The so--called Factorization Method plays an important role
in the search for  quantum systems for which the spectrum
of the Hamiltonian operators is completely known.\cite{IH,GK85,CarMarPerRan}
It is closely related with the existence of an 
intertwining operator,\cite{FerHusMiel98,FerHus99} with Supersymmetric
Quantum Mechanics\cite{CoopKhaSuk95} and 
Darboux transformations in this last context.\cite{MatSal91}
Moreover, these techniques have important generalizations to higher 
dimensional spaces,\cite{AndBorIofEid84}
to higher order factorization 
operators,\cite{aicd95,bs95,Fer97,AndCanIofNis00}
and to the class of systems with partial algebraization 
of the spectrum,\cite{Tur88,Shi89,ShiTur89_1}
among others. Actually, most exactly solvable potentials can 
be obtained by making use of an appropriate intertwining operator 
transformation. 

Looking for a factorization of a given Hamiltonian amounts to find a 
constant $d$ and a solution of a Riccati differential equation for the
superpotential function. Once a solution has been found,
a partner potential is defined such that it has almost the same spectrum 
as the original one. However, the point is that with a different choice
for the solution of the mentioned Riccati equation, 
a different partner potential is obtained. This fact has been shown 
to be very useful for the search of isospectral potentials, an idea
due to Mielnik\cite{Mie84} and later developed in other 
articles.\cite{Fer84,Nie84,Don87,DiaNegNieRos99} 

Moreover, the ambiguity in the definition of \emph{a partner} potential
of a given one is twofold: firstly, due to the choice of the factorization 
energy $d$, which is not unique in general, and then, it arises
the ambiguity in the election of the solution of the corresponding 
Riccati equation. These two ambiguities are more or less known and
are implicitly used or mentioned in many papers. However, we feel that 
it is worth having a new look at the subject in its own right because 
its understanding allows to interpret certain facts treated in the literature  
as consequences of this undetermination. In addition, these 
ideas have a great influence in the same concept of partnership when 
applied to the subclass of Shape Invariant partner potentials.

\emph{Shape Invariance} is an important concept in the theory of exactly 
solvable systems, which was explicitly introduced
by Gendenshte\"{\i}n,\cite{Gen} although the basic idea
was already present, to some extent, in the classic work of 
Infeld and Hull.\cite{IH} That was suggested by some authors several 
years ago\cite{Sta2,MonSal} and has been shown
with some detail recently.\cite{CarRam2} The ambiguity in the definition of 
the partner potential is inherited in the case of Shape Invariance,
so one may wonder to what extent it makes sense the relation 
between a potential and \emph{its} partner characterizing such a kind
of problems.

Therefore, two main questions arise. Are there different solutions 
for the same Riccati equation leading to the same partner?. On the other
hand, if the Shape Invariance condition holds for a certain  partner,
is it also true for any other possible partner?. 
Our aim is to analyze these questions using, among other things, 
the machinery for dealing with Riccati equations developed in 
previous papers.\cite{CarRam2,CarRam} 

The letter is organized as follows.
In the next section we will review briefly the concepts of partner potential, 
when \emph{one} potential is given, and Shape Invariance. In Section 3 
we look at the abovementioned ambiguities.
In two recent papers \cite{FerNegOlm96,DiaNegNieRos99} two alternative 
factorizations for different choices of the constant $d$ are given.
We will analyze this point in Section 4, showing that 
the ambiguity in the factorization energy, and hence the existence of some
factorizations is due, in turn, to certain parameter invariance
symmetry of the given potential. Moreover, if the given potential can be 
considered as being part of a pair of Shape Invariant partner potentials,
more factorizations can be found.
We develop these ideas with some illustrative examples. 

\section{Factorization method and shape invariance}

The simplest way of generating a new exactly solvable Hamiltonian
$\wt H$ from a known one $H$ is
just to consider an invertible bounded operator $B$, with bounded inverse,
and defining $\widetilde H=BHB^{-1}$. This new Hamiltonian $\wt H$
has the same spectrum as the starting Hamiltonian $H$. 
As a generalization,\cite{CarMarPerRan}
we will say that two Hamiltonian operators $H$ and $\wt H$ are $A$--related
when  $AH=\wt HA$, where  $A$ may  be singular. In this case,
if $\psi $ is an eigenvector
of $H$ corresponding to the eigenvalue $E$ and $A\psi\not =0$, then, at least
formally,  $A\psi$ is also an eigenvector of $\wt H$ corresponding to
the same eigenvalue $E$.

If we assume that the intertwining operator $A$ 
is a first order differential operator,
$$A=\frac {d}{dx}+W(x)\ ,\quad\quad\mbox{and}\quad\quad 
A^{\dagger}=-\frac{d}{dx}+W(x)\ ,
$$
then the relation $AH=\wt HA$, with
\begin{equation}
H=-\frac{d^2}{dx^2}+V(x)\,,\qquad \wt H=-\frac{d^2}{dx^2}+\wt V(x)\ ,
\label{defHHtil}
\end{equation}
leads to
$$V=-2W'+\wt V, \qquad W(V-\wt V)=-W''-V'\ ,
$$
and taking into account the first equation, the second one becomes
$2WW'=W''+V'$, which can easily be integrated giving
\begin{eqnarray}
V&=&W^2-W'+d\,,         \label{ricV} \\ 
\wt V&=&W^2+W'+d\,,     \label{ricVtil}
\end{eqnarray}
where $d$ is an integration constant. The important point here is that 
$H$ and $\wt H$, given by (\ref{defHHtil}), can be related by
a first order differential operator $A$ of the form given above if, and only
if, there exists a constant $d$ and a function $W$ such that the pair
of Riccati equations (\ref{ricV}) and (\ref{ricVtil}) are 
satisfied \emph{simultaneously}. 
Moreover, this means that both Hamiltonians can be factorized as 
\begin{equation}
H=A^{\dag}A+d\,,\qquad \wt H=AA^{\dag}+d\ .\label{factorHHtil}
\end{equation}

Adding and subtracting equations (\ref{ricV}) and (\ref{ricVtil})
we obtain the equivalent pair which relates $V$ and $\wt V$
\begin{eqnarray}
\wt V-d&=&-(V-d)+2W^2\,,       \label{relVVtilcuad} \\
\wt V&=&V+2 W'\,.              \label{relVVtilder} 
\end{eqnarray}
The function $W$ satisfying these equations is usually called 
\emph{superpotential}, the constant $d$ \emph{factorization energy} or
\emph{factorization constant} and  
$\wt V$ and $V$ (resp. $\wt H$ and $H$) are said to be
\emph{partner} potentials (resp. Hamiltonians). 

 Gendenshte\"{\i}n 
took equations (\ref{ricV}) and (\ref{ricVtil}) 
as a definition of the functions $V$, $\wt V$ in terms of the function $W$ and
some constant $d$. After, he supposed that $W$ did depend on certain
set of parameters $a$, $W=W(x,a)$, and as a consequence 
$V=V(x,a)$ and $\wt V=\wt V(x,a)$ as well.
Then, the necessary condition for $\wt V(x,a)$  to be  of the
same form as $V(x,a)$, maybe for a different choice 
of the values of the parameters involved in $V$,
is  known as Shape Invariance. 
More explicitly, it amounts to assume the further relation
between $V(x,a)$ and $\widetilde V(x,a)$
\begin{equation}
\widetilde V(x,a)=V(x,f(a))+R(f(a))\,,\label{SIcond}
\end{equation}
where $f$ is an (invertible) transformation on the 
parameter space $a$ and $R$ is some function. 
The main advantage of these problems is that the 
complete spectrum of the corresponding Hamiltonians $H$ and $\wt H$ can
be found easily.\cite{Gen}
Let us remark that it is  the choice of $a$ 
and $f(a)$ what defines the different classes of Shape Invariant
potentials. The function $f$ may be even 
the identity, $f(a)=a$.\cite{AndCanIofNis00}
For a more detailed information see, for example, Ref. 24 .
%\cite{CarRam2}.

\setcounter{footnote}{0}
\renewcommand{\thefootnote}{\alph{footnote}}

\section{On the ambiguity in the definition of the partner potential\label{partner}}
\noindent

Given \emph{one} potential function $V$, the equation (\ref{ricV}) 
to be solved when searching for a superpotential function $W$, once $d$ is fixed,
is a Riccati equation. In general, its general solution cannot be found by means 
of quadratures. However, now
we only need to compare solutions of the same equation when a particular solution
is known. In such a case, it is well known that 
its general solution can be written using 
two quadratures. For a group theoretical explanation of this fact, 
see, for example, Ref. 25. %\cite{CarRam}.
Our aim now is to study a bit further the general solution of (\ref{ricV})
in that situation and analyze 
the corresponding possible partner potentials $\wt V$.

It is well known\cite{CarRam2,CarRam} that if  $W_p$ is a 
particular solution of (\ref{ricV}) for some specific 
constant $d$, the change of variable 
\begin{equation}
v=\frac{1}{W_p-W}\,,\quad\mbox{with inverse} 
\quad W=W_p-\frac{1}{v}\,,                      \label{ch_1sol_usual}
\end{equation}
 transforms 
(\ref{ricV}) into  the inhomogeneous first order linear equation for $v$
\begin{equation}
\frac{dv}{dx}=-2\,W_p\,v+1\,, \label{vwp}
\end{equation}
which has the general solution
\ba
v(x)=\frac{\int^x \exp\big\{2\int^\xi W_p(\eta)\,d\eta\big\}\,d\xi+F}
{\exp\big\{2\int^x W_p(\xi)\,d\xi \big\}}\,,
\label{gener_v}
\ea
where $F$ is an integration constant. 
Therefore, the general solution of (\ref{ricV}) reads as  
\ba
W_g(x)=W_p(x)-\frac{\exp\big\{2\int^x W_p(\xi)\,d\xi \big\}}
{\int^x \exp\big\{2\int^\xi W_p(\eta)\,d\eta\big\}\,d\xi+F}\,.
\label{gsW1}
\ea

We will review now the concept of partnership given \emph{one} potential $V$.
We have to find a constant $d$ and at least one particular solution $W_p$ of the
Riccati equation (\ref{ricV}). Then, \lq\lq the partner\rq\rq\  $\wt V$ 
is constructed by using (\ref{ricVtil}) or equivalently (\ref{relVVtilder}).
But these formulas explicitly show that $\wt V$ does depend
upon the choice of the particular solution of (\ref{ricV}) considered.
Since the general solution of (\ref{ricV}) can be written as
$W_g=W_p-1/v$, where $v$ is given by (\ref{gener_v}),
the general solution obtained for $\wt V_g$ is, according to (\ref{relVVtilder}), 
\begin{equation}
\wt V_g=\wt V_p-2\frac d{dx}\left(\frac 1v\right)\ .\label{expwtg}
\end{equation} 
 
This answers one of the questions in the introduction: all the 
partner potentials, obtained by using (\ref{expwtg}) are different, 
apart from  the trivial case in which $W_p$ and $V$ are constant, 
because the differential equation (\ref{vwp}) only admits a constant solution 
when $W_p$ is constant. 

This implies that 
\lq\lq the partner\rq\rq\ of \emph{one} given potential is not a well 
defined concept and it seems better 
to say that an ordered pair $(V,\wt V)$ is a supersymmetric pair of 
partner potentials if there exists a constant $d$ and a function $W$ such that 
this last 
is a common solution of the Riccati equations (\ref{ricV}) and (\ref{ricVtil})
constructed with these potentials, respectively. 
Of course the preceding comment shows that in such a case the 
superpotential function $W$ is essentially unique for each $d$,
which moreover makes the problem of $A$--related Hamiltonians be 
well  defined. Note as well that this reformulation of partnership
comprehends the situation where $V$ is the potential we have started 
this section with, $\wt V$ is one of the functions obtained from
(\ref{expwtg}) for a \emph{specific} value of the constant $F$, and $W$
is obtained from (\ref{gsW1}) for \emph{the same} value of $F$.  

Now we will show what consequences have this undetermination in the 
subclass of Shape Invariant potentials.
For that, we should use instead of (\ref{ricV}) 
and (\ref{ricVtil}) the equations 
\ba
V(x,a)-d(a)&=&W^2(x,a)-W'(x,a)\,,               \label{ricVSI}  \\
\widetilde V(x,a)-d(a)&=&W^2(x,a)+W'(x,a)\,,    \label{ricVtilSI}
\ea
where now the factorization constant depends on the parameter $a$
(see Ref. 24, Sec. 3 for details). %\cite{CarRam2}
Consider a particular solution $W_p(x,a)$ of equation (\ref{ricVSI})
for some specific constant $d(a)$, such that it is also a particular 
solution of (\ref{ricVtilSI}), being $V(x,a)$ and $\widetilde V(x,a)$
related by the further condition (\ref{SIcond}). As in the previous case,
we can consider the general solution of (\ref{ricVSI}) starting from 
$W_p(x,a)$, which is 
\ba
W_g(x,a,F)=W_p(x,a)+g(x,a,F)\,,
\label{gsW1_SI}
\ea
where $g(x,a,F)$ is defined by
\ba
g(x,a,F)=-\frac{\exp\big\{2\int^x W_p(\xi,a)\,d\xi \big\}}
{\int^x \exp\big\{2\int^\xi W_p(\eta,a)\,d\eta\big\}\,d\xi+F}\,,
\label{def_g}
\ea
being $F$ a integration constant. Note that the particular 
solution $W_p(x,a)$ is obtained from (\ref{gsW1_SI}) as $F\rightarrow \infty$.
Then, inserting $W_g(x,a,F)$
into an equation like (\ref{ricVtilSI}) we obtain the general family
of partner potentials 
\begin{equation}
\widetilde V(x,a,F)=\widetilde V(x,a)-2 g^\prime(x,a,F)\,.
\label{pot_gen_SI}
\end{equation}

The question now is whether the condition (\ref{SIcond}) is maintained
when we consider $\widetilde V(x,a,F)$ and $V(x,a)$ instead of 
$\widetilde V(x,a)$ and $V(x,a)$. Then, we ask for
\begin{equation}
\widetilde V(x,a,F)=V(x,f(a))+\overline R(f(a),F)\,,
\label{SIcond_gen}
\end{equation}
for some suitable $F$, where $f$ is the same as in (\ref{SIcond}), and
$\overline R(f(a),F)$ is a number not depending on $x$, maybe different
from the $R(f(a))$ of (\ref{SIcond}). Taking into account (\ref{SIcond}) and
(\ref{pot_gen_SI}), the equation (\ref{SIcond_gen}) reads as 
$$
2 g^\prime(x,a,F)=\overline R(f(a),F)-R(f(a))\,,
$$
that is, $2 g^\prime(x,a,F)$ should be a constant, which we name as 
$k$ for brevity. Integrating respect to $x$ we obtain 
$2 g(x,a,F)=k x+l$, being $l$ another constant depending at most on $a$ and $F$.
Since $g(x,a,F)$ is, on the other hand, given by (\ref{def_g}), it follows
\begin{equation}
\int^x \exp\big\{2\int^\xi W_p(\eta,a)\,d\eta\big\}\,d\xi+F
=-\frac{2\,\exp\big\{2\int^x W_p(\xi,a)\,d\xi \big\}}{k x+l}\,.
\label{eq_interm}
\end{equation}
Differentiating this last equation, and solving for $W_p(x,a)$ we obtain
$$
W_p(x,k,l)=\frac 1 4 \bigg(\frac{2 k}{k x+l}-(k x+l)\bigg)\,,
$$
where we have made explicit that the 
parameter space should be $a=\{k,l\}$. Introducing this
expression in (\ref{eq_interm}) and performing the integrations, we obtain
$$
-2\,e^{-x(k x+2 l)/4}+F=-2\,e^{-x(k x+2 l)/4}
$$
and hence $F=0$. Now we have to check whether this particular 
case we have found, which is the only candidate for 
fulfilling (\ref{SIcond_gen}), satisfy our hypothesis (\ref{SIcond}).
The partner potentials defined by $W_p(x,k,l)$ and equations
(\ref{ricVSI}), (\ref{ricVtilSI}) are
\ba
&&V(x,k,l)-d(k,l)=W_p^2(x,k,l)-W_p^\prime(x,k,l)
=\frac{(k x+l)^2}{16}+\frac{3\,k^2}{4\,(k x+l)^2}\,,	\nonumber\\
&&\widetilde V(x,k,l)-d(k,l)=W_p^2(x,k,l)+W_p^\prime(x,k,l)
=\frac{(k x+l)^2}{16}-\frac{k^2}{4\,(k x+l)^2}-\frac k 2\,. \nonumber	
\ea
Now, we have to find out whether there are some transformation of the
parameters $\{k,l\}$ such that the condition (\ref{SIcond}) be satisfied.
Denoting the transformed parameters as $\{k_1,l_1\}$ for simplicity,
we have
\ba
&&\widetilde V(x,k,l)-V(x,k_1,l_1)=d(k,l)-d(k_1,l_1)-\frac k 2 	
-\frac 1 4 \bigg(\frac{3\, k_1^2}{(k_1 x+l_1)^2}
+\frac{k^2}{(k x+l)^2}\bigg)					\nonumber\\
&&\quad\quad\quad+\frac 1 {16}((k-k_1)x+l-l_1)((k+k_1)x+l+l_1)\,.\nonumber
\ea
The right hand side of this equation must be a constant and therefore,
each of the different dependences on $x$ must vanish. 
The term $((k-k_1)x+l-l_1)((k+k_1)x+l+l_1)$ vanish for the
combinations $k_1=-k,\,l_1=-l$ or $k_1=k,\,l_1=l$, apart form the
case $k_1=-k_1=k=0$, which will be studied separately. However,
the term 
$$
\frac{3\, k_1^2}{(k_1 x+l_1)^2}+\frac{k^2}{(k x+l)^2}
$$ 
is equal to $4\,k^2/(k x+l)^2$ for both combinations and does not vanish. 
Then, the Shape Invariance hypothesis is not satisfied. In the case of 
$k=0$ we have that the corresponding $W_p(x,a)$ is a constant and hence
provides the trivial case where the corresponding partner 
potentials are constant as well.

This answers the other question of the introduction, and it is 
closely related with the previous one. That is, if the Shape Invariance
condition holds for a possible partner, then it does not hold for any other
choice of partner, apart from the trivial case where all the involved functions
are constant.

As a consequence of all the previous, it would be better to 
reformulate the Shape Invariance condition (\ref{SIcond}) in terms of 
appropriate $W$ and $d$ only.
Now, considering a particular
common solution $W(x,a)$ of (\ref{ricVSI}) and (\ref{ricVtilSI}) for some 
$d(a)$, jointly with (\ref{SIcond}) allows to write this last condition as 
\begin{equation}
W^2(x,a)-W^2(x,f(a))+W'(x,f(a))+W'(x,a)=R(f(a))\ ,\label{SIsp}
\end{equation}
where $R(f(a))=d(f(a))-d(a)$. This way, beginning from $W(x,a)$ and $d(a)$
which solve (\ref{SIsp}) for some $f$, we will obtain through (\ref{ricVSI}) and 
(\ref{ricVtilSI}) well defined Shape Invariant partner potentials 
$(V(x,a),\wt V(x,a))$ by construction. 
In the celebrated article by Infeld and Hull\cite{IH} 
the key point is indeed to solve an equation of type (\ref{SIsp}) 
(see their equation (3.1.2)). Similarly, in Ref. 27, %\cite{CoopGinKha87},
Sec. VI, the main point in the classification of Shape Invariant
potentials they propose is to find solutions of an equation of type
(\ref{SIsp}) (see their formula (2.22)). An equation of type 
(\ref{SIsp}) also plays a central role in more recent 
articles,\cite{CarRam2,CarRam3}  
where some of the results of Infeld and Hull\cite{IH}
are reviewed and put in connection with the concept
of Shape Invariance, and 
Shape Invariant partner potentials depending on $n$ parameters transformed
by translation are found, giving a solution to a previously unsolved
problem.\cite{CoopGinKha87}
However, it seems that a justification of why it is necessary 
to solve an equation of type (\ref{SIsp}) when searching for 
well defined Shape Invariant partner potentials has not been given 
explicitly in the literature up to now.    

\section{Parameter invariance and Shape Invariance: existence of several 
factorizations\label{SIPI}}
\noindent

We will analyze in this section what happens if there 
exists a transformation in the parameter space, 
$g:a\mapsto g(a)$ such that leaves the potential $V(x,a)$ in
(\ref{ricVSI}) invariant.
Then, whenever $(W(x,a),\,d(a))$ is a solution of (\ref{ricVSI}),
we will have another different solution provided $W(x,g(a))\neq W(x,a)$.
In fact, if we transform all instances of $a$ in 
(\ref{ricVSI}) by the map $g$, and use such an invariance property, 
it follows that we have another solution $(W(x,g(a)),\,d(g(a)))$ 
of (\ref{ricVSI}) in addition to $(W(x,a),\,d(a))$.
Inserting each of these pairs
into (\ref{ricVtilSI}) we will obtain in general different partner
potentials $\wt V(x,g(a))$ and $\wt V(x,a)$ of $V(x,a)$.
This also gives an example of the fact that there may exist several different 
constants $d$ such that we could find a particular solution $W$
of an equation of type (\ref{ricV}) or (\ref{ricVSI}) for a fixed $V$.

Another interesting case in which new factorizations can be generated from known
ones is when we have a pair of partner potentials $V(x,a)$ and 
$\widetilde V(x,a)$ satisfying  the  Shape Invariance condition (\ref{SIcond}),
properly understood. In this case this condition shows that 
$$
V(x,a)=\widetilde V(x,f^{-1}(a))-R(a)\,,
$$
or in terms of the Hamiltonians,
$$
H(a)=\widetilde H(f^{-1}(a))-R(a)\,,
$$
which provides an alternative  factorization
for $H(a)$:
$$
H(a)=\left(\frac{d}{dx}+W(x,f^{-1}(a))\right)
\left(-\frac{d}{dx}+W(x,f^{-1}(a))\right)+d(f^{-1}(a))-R(a)\,,
$$
where it has been used (\ref{factorHHtil}) with $A(a)=\frac{d}{dx}+W(x,a)$ and 
$A^{\dagger}(a)=-\frac{d}{dx}+W(x,a)$. So, had we started \emph{only} 
with the potential $V(x,a)$ of this paragraph, we would have been able to 
find a factorization of $H(a)$ as a product of 
type $A^{\dagger}(a)A(a)+\mbox{Const.}$ and another as a product 
$A(f^{-1}(a))A^{\dagger}(f^{-1}(a))+\mbox{Const.}$, 
being these last constants different in general.

Of course one could have both situations of the preceding paragraphs 
at the same time. We shall illustrate them in the next subsection.

\subsection{Illustrative examples\label{examples}}
\noindent
As a first example we will explain 
the {\sl four--way factorization} of the isotropic harmonic oscillator,
introduced in Ref. 26, %\cite{FerNegOlm96}, 
pp. 388--9.
In their notation, the potential 
and Hamiltonian of interest are
$$
V(r,l)=\frac{l(l+1)}{r^2}+r^2\,,\quad\quad\quad H(l)=-\frac{d^2}{dr^2}+V(r,l)\,,
$$
where the independent variable is $r\in (0,\infty)$ and the set of parameters 
is simply $l$. Their factorization (6) is 
\ba
H(l)=\left(-\frac{d}{dr}+\frac{l}{r}+r\right)
\left(\frac{d}{dr}+\frac{l}{r}+r\right)-(2 l-1)\,,
\label{Hl}
\ea
from where it is suggested that $W(r,l)=\frac{l}{r}+r$. Substituting it in
$V(r,l)=W^2(r,l)-\frac{W(r,l)}{dr}+d(l)$ we obtain $d(l)=-(2 l-1)$, so 
(\ref{Hl}) is the appropriate version of our (\ref{factorHHtil}) as expected.
Now, as the potential $V(r,l)$ is invariant under the map $g:l\mapsto -l-1$,
we will obtain a new solution $(W(r,g(l))\,,d(g(l)))=(W(r,-l-1)\,,d(-l-1))$ 
of the equation 
$$
V(r,l)=W^2-\frac{dW}{dr}+d\,.
$$  
But $W(r,g(l))=W(r,-l-1)=-\frac{l+1}{r}+r$ and $d(g(l))=d(-l-1)=2l+3$, which
is exactly what corresponds to the factorization (4) of Ref. 26.
%\cite{FerNegOlm96}, page 389. 
The factorizations (5) and (7) {\it loc. cit.} are related in a similar way;
(7) is obtained from (5) by means of the change $g:l\mapsto -l-1$ as well.
 
As far as the relation between their factorizations (6) and (5) 
{\it loc. cit.} is concerned, we have already seen that, 
from their factorization (6), here reproduced as
(\ref{Hl}), it follows $W(r,l)=\frac{l}{r}+r$, and thus, the corresponding 
$\widetilde V(r,l)$ through (\ref{ricVtilSI}) is
$$
\widetilde V(r,l)=W^2(r,l)+\frac{d W(r,l)}{dr}+d(l)=\frac{l(l-1)}{r^2}+r^2+2\,.
$$
Then it is very easy to check that $\widetilde V(r,l)=V(r,f(l))+R(f(l))$, where 
$R(l)=2$ for all $l$, and $f$ is defined either by $f(l)=l-1$ or $f(l)=-l$. 
We obtain 
$$
H(l)=\widetilde H(l+1)-R(l)\,,\quad\quad\quad V(r,l)=\widetilde V(r,l+1)-R(l)\,,
$$ 
and  
$$
H(l)=\widetilde H(-l)-R(l)\,,\quad\quad\quad V(r,l)=\widetilde V(r,-l)-R(l)\,,
$$ 
as well. In this way the factorization (5) of Ref. 26 %\cite{FerNegOlm96} 
is achieved.

As a second example we will consider the modified P\"oschl--Teller 
potential, analyzed in an interesting recent article.\cite{DiaNegNieRos99} 
The potential is now
\ba
V(x,\alpha,\lambda)=-\alpha^2 \frac{\lambda(\lambda-1)}{\cosh^2 \alpha x}\,,
\ea
where $x\in(-\infty,\infty)$ and  
$\alpha>0$, $\lambda>1$ are two real parameters.

Two different particular solutions 
$(W(x,\alpha,\lambda),d(\alpha,\lambda))$ of the Riccati equation
$$
W^2-W^\prime=V(x,\alpha,\lambda)-d\,,
$$ 
have been found in Ref. 20, %\cite{DiaNegNieRos99}, 
p. 8450, namely,
\ba
(W_1(x,\alpha,\lambda),d_1(\alpha,\lambda))
&=&(-\lambda\,\alpha \tanh^2 \alpha x,-\lambda^2 \alpha^2)\,,           \nonumber\\
(W_2(x,\alpha,\lambda),d_2(\alpha,\lambda))
&=&(-(1-\lambda)\,\alpha \tanh^2 \alpha x,-(1-\lambda)^2\alpha^2)\,.    \nonumber
\ea
It is clear that the second pair is obtained from the first by means of the
parameter transformation $g:(\alpha,\lambda)\mapsto(\alpha,1-\lambda)$. 
The reason is that $V(x,\alpha,\lambda)$ is invariant under $g$, 
or more precisely, its factor $\lambda(1-\lambda)$.

The associated partner potentials $\widetilde V(x,\alpha,\lambda)$ obtained using 
(\ref{ricVtilSI}), are 
\ba
\widetilde V_1(x,\alpha,\lambda)
&=&W_1^2(x,\alpha,\lambda)+W_1^\prime(x,\alpha,\lambda)+d_1(\alpha,\lambda)
=-\alpha^2 \frac{\lambda(\lambda+1)}{\cosh^2 \alpha x}\,,       \nonumber\\
\widetilde V_2(x,\alpha,\lambda)
&=&W_2^2(x,\alpha,\lambda)+W_2^\prime(x,\alpha,\lambda)+d_2(\alpha,\lambda)
=-\alpha^2 \frac{(\lambda-1)(\lambda-2)}{\cosh^2 \alpha x}\,.   \nonumber
\ea
We see that both of the previous functions are just second degree
monic polynomials in $\lambda$, with roots spaced one unit, 
times $-\alpha^2/\cosh^2 \alpha x$,
like $V(x,\alpha,\lambda)$ itself. It is then obvious that a 
translation of the type $\lambda\mapsto \lambda-b$ or 
$\lambda\mapsto c-\lambda$ should transform  
$\widetilde V_1(x,\alpha,\lambda)$ and $\widetilde V_2(x,\alpha,\lambda)$ into 
$V(x,\alpha,\lambda)$. This is in fact so, since 
$V(x,\alpha,\lambda)=\widetilde V_1(x,\alpha,f^{-1}(\lambda))$, where
$f$ is defined either by $f(\lambda)=\lambda-1$ or $f(\lambda)=-\lambda$,
and similarly $V(x,\alpha,\lambda)=\widetilde V_2(x,\alpha,f^{-1}(\lambda))$ when 
$f(\lambda)=\lambda-1$ or $f(\lambda)=2-\lambda$. 

In this way one could propose other different factorizations for the potential 
$V(x,\alpha,\lambda)$, being able in principle to 
make a differential operator analysis for this potential similar 
to what it is done in Ref. 26 %\cite{FerNegOlm96} 
for the first example of this Subsection.

%\vfill\eject

\nonumsection{Acknowledgments}
\noindent
A. R. thanks the Spanish Ministerio de 
Educaci\'on y Cultura for a FPI grant, research project 
PB96--0717. Support of the Spanish DGES (PB96--0717) is also acknowledged. 

\nonumsection{References}

\end{document}